\documentclass[twocolumn,showpacs,amsmath,aps]{revtex4}
\usepackage{txfonts}
\usepackage{graphicx,amssymb,mathrsfs,bm,color,times}

\newcommand{\be}{\begin{equation}}
\newcommand{\ee}{\end{equation}}
\newcommand{\bea}{\begin{eqnarray}}
\newcommand{\eea}{\end{eqnarray}}
\newcommand{\bsube}{\begin{subequations}}
\newcommand{\esube}{\end{subequations}}

\newcommand{\Eq}[1]{Eq.\,(\ref{#1})}

\newcommand{\dg}{\dagger}
\makeatletter 
    
    \newcommand{\Rmnum}[1]{\expandafter\@slowromancap\romannumeral #1@}
\makeatother

\begin{document}
\title {Dynamic Quantum Tunneling in Mesoscopic Driven Duffing Oscillators}

\author{Lingzhen Guo$^{1,2}$, Zhigang Zheng$^{1}$, Xin-Qi Li$^{1}$\footnote{lixinqi@bnu.edu.cn} and YiJing Yan$^3$}
\author{}
\affiliation{ $^1$\mbox{Department of Physics, Beijing Normal
University, Beijing 100875, China}\\ $^2$\mbox{Institut f\"ur
Theoretische Festk\"orperphysik, Karlsruhe Institute of
Technology, 76128 Karlsruhe, Germany}\\ $^3$\mbox{Department of
Chemistry, Hong Kong University of Science and Technology,
Kowloon, Hong Kong}}

\date{\today}

\begin{abstract}
We investigate the dynamic quantum tunneling between two attractors
of a mesoscopic driven Duffing oscillator.
We find that, in addition to inducing remarkable quantum shift of
the bifurcation point, the {\it mesoscopic} nature also results in
a perfect linear scaling behavior for the tunneling rate with
the driving distance to the shifted bifurcation point.
\end{abstract}

 \maketitle


\section{Introduction}

In the light of nanomechanics \cite{Koz07}
and as a qubit readout device for superconducting qubits
(i.e., the Josephson bifurcation amplifier)
\cite{Siddiqi207002,Siddiqi027005,Devoret014524,Mooij119},
the quantum dynamical properties of the driven Duffing oscillator (DDO)
gained renewed interest in the past years
\cite{Katz07,Pea06,Dyk05,Dykman042108,Dykman011101,MDQT,low-lying},
since these real systems can be well described by a DDO
under proper parameter conditions.
While some experimental results can be understood
by classical means \cite{Gro04,Gro05,Cas07,Gro10},
the success of such interpretations
is basically due to the fact that the quantum levels are hardly
distinguishable from classical nonlinear excitations \cite{Gro05},
and the data were classical in origin \cite{Gro10}.
Actually, possible quantum behaviors and the typical features
of the DDO are of great interest to the community mentioned above.
For instance, in quasiclassical limit,
the quantum feature in the bistable region is investigated
against the conventional classical DDO,
by simulating a Lindblad-type master
equation and comparing the Wigner function with classical
probability distribution in phase space \cite{Katz07}.
Interesting result reached there reveals that the quantum effect
makes easier the transition between the bistable states.
Moreover, in Refs.\ \cite{Dyk05,Dykman042108,Dykman011101,MDQT},
the switching rate between the bistable states near the bifurcation point,
owing to quantum and/or thermal fluctuations,
were carried out by means of the WKB theory or semiclassical methods
such as the mean-first-passage-time approach.

While much more work should be done for unambiguous
demonstrations of the quantum behaviors of the DDO,
studies of the quantum dynamics of DDO
in certain fully quantum regimes are valuable and necessary
at the early stage before attacking the whole difficult problem.
Indeed, in a fully quantum regime which only involves a few
quantum levels of the potential well into the dynamics,
quantum behaviors of the DDO such as resonant tunneling and
photon-assisted tunneling are discussed in terms of
amplitude and phase responses to the driving frequency \cite{Pea06}.
In the present work, we consider further an {\it intermediate}
quantum regime, say, the quantum dynamics of  a {\it mesoscopic} DDO,
which involves more than ten quantum levels in the nonlinear dynamics.
Obviously, this is a regime between the quantum few-level and
the classical dense-level (or continuum) limits \cite{note-1},
where the quantum effect should be crucially important,
and at the same time the DDO's nonlinear characteristics can emerge as well.
More specifically, we will focus on the {\it dynamic quantum tunneling}
between the two attractors in the bistable region \cite{note-2},
in particular accounting for
a quantum shift of the bifurcation point associated with the
{\it mesoscopic} nature, and extracting a new scaling
exponent with the driving distance to the shifted bifurcation point.


\section{Model and Qualitative Considerations}

The DDO is typically described by
\begin{equation}\label{DDO}
H_S(t)=p^2/2m+m\Omega^2 x^2/2 - \gamma x^4+F(t)x,
\end{equation}
where $F(t)=2F_{0}\cos(\nu t)$ stands for the external driving force.
Moreover, a DDO is inevitably influenced by its surrounding environment,
which can be quantum mechanically modeled by a set of harmonic oscillators,
$H_E=\sum_{i}(m_i\omega_i^2 x_i^2/2+ p_i^2/2m_i)$.
In a spirit of the Feynman-Vernon-Caldeira-Legget model \cite{Wei92},
the coupling between the Duffing oscillator and the environment
is well described by $H_I=- x\sum_{i}\lambda_i x_i $.
After introducing the spectral density function,
$J(\omega)=\pi\sum_{i}\lambda_{i}^2\delta(\omega-\omega_{i})
/(2m_i\omega_i)$, an important situation is the so-called Ohmic case,
corresponding to $J(\omega)=m\kappa\omega\mathrm{exp}(-\omega/\omega_c)$,
where $\kappa$ is the friction coefficient,
and $\omega_{c}$ a high frequency cutoff.
In the Ohmic case, one can successfully recover the classical
friction force \cite{Wei92},
which plays an essential role in the
classical dynamics of DDO \cite{Devoret014524}.
In this work, we will employ a master equation
approach to simulate the quantum dynamics of DDO.
This treatment is equivalent to the (quantum) Langevin equation
in which not only the {\it friction} but also the {\it fluctuations}
are taken into account.
For the convenience of later use, we also introduce here the operator
$b=\sum_i\lambda_i b_i/{\sqrt{2}}$, with
$b_i=(m_i\omega_i x_i+i p_i)/{\sqrt{2m_i\hbar\omega_i}}$.

Notice that the Duffing oscillator described by \Eq{DDO}
has only finite number of bound states.
In the absence of external driving, this becomes clear from the so-called
{\it soft nonlinearity} ($\gamma>0$) potential profile \cite{Devoret014524},
$V(x)=\frac{m\Omega^2}{2}x^2-\gamma x^4$,
which defines a single well with identical barrier height
$V_0=m^2\Omega^4/(16\gamma)$ at $x=\pm\sqrt{m\Omega^2/(4\gamma)}$.
Nevertheless, compared to the {\it hard nonlinearity} ($\gamma < 0$)
potential well \cite{Pea06}, which supports infinite number
of bound states, the resultant nonlinear dynamics
has no fundamental differences.
This is because we are interested in a region not too far
from the bottom of the anharmonic potential well.
For the soft nonlinearity potential well, the chosen parameters can
assure that the unbounded states have negligible effect on the dynamics.
This also indicates that the ``metastability"
to be addressed in the following
has no relevance to the ``unbounded"
states at very high excitations in the soft well.
As a rough estimate, the number of bound states is the ratio of
$V_0$ and $\hbar\Omega$, which gives
$ N=\frac{m^2\Omega^4}{16\gamma
\hbar\Omega}=\frac{m\Omega}{16\hbar\tilde{\gamma}}
=\frac{\aleph}{16\tilde{\gamma}} $.
Here we introduced $\aleph\equiv m\Omega/\hbar$,
and $\tilde{\gamma}\equiv \gamma/(m\Omega^2)$.
In our model, $\gamma=m\Omega^2/24$,
so approximately the number of bound state is $3\aleph/2$.
In the experiment of Ref.\ \cite{Siddiqi207002}, for instance,
a rough estimate gives $\aleph \simeq 366$.
In the present work, however, we will set $\aleph\simeq 12$ by altering
the circuit parameters.
This defines a {\it mesoscopic} regime  for the DDO
with 18 quantum levels involved in the dynamics.
Also, we assume an accessible temperature of $5$ mK.

Now we present a qualitative analysis to the driving dynamics of DDO,
respectively, in the laboratory frame and in a rotating frame.
In Fig.\ 1(a) we show the energy level diagram of the Duffing oscillator
in the absence of driving.
To the second-order perturbation of the quartic potential,
the energy level reads
$E_n=[n+1/2-3\tilde{\gamma}(2n^2+2n+1)/(4\aleph)]\hbar\Omega$.
Accordingly, the adjacent level spacing,
$\Delta E_n=E_n-E_{n-1}=(1-3\tilde{\gamma}n/\aleph)\hbar\Omega$,
decreases with $n$.
This property, together with a negative frequency detuning
(e.g. $\delta=1-{\nu}/{\Omega}=0.065$ in later simulation),
would result in the  bistability behavior.
Qualitatively speaking, for weak driving, the oscillator will largely remain
in the initial ground state; for stronger driving, however,
it will be increasingly excited to high energy states around $n^*$,
roughly determined by
$\hbar\nu=\Delta E_{n^*}$, leading to $n^*=\aleph \delta/(3\tilde{\gamma})$.

We notice that an alternative (a better) way to understand the quantum
dynamics is in a rotating frame with the driving frequency $\nu$,
where the driving field becomes time independent.
This can be implemented by applying a rotating transformation
$ {U}(t)=\mathrm{exp}\{-i\nu t {a}^\dagger {a}\}$,
where $a \ (a^{\dagger})$ is the annihilation (creation) operator of the DDO.
Dropping fast rotating terms, i.e., under the rotating wave approximation (RWA),
we obtain
$ {\tilde{H}_S}
=\delta [ p^2/(2m)+ (1/2) m\Omega^2{x}^2 ]
-6\gamma/(4m^2\Omega^4)
[ {p}^2/(2m) + (1/2) m\Omega^2 {x}^2 ]^2 + F_0 {x}. $
In the absence of driving, this rotated Hamiltonian is
naturally diagonal, with the eigenstates of harmonic oscillator ${\psi}_n$,
and eigenvalues
$\tilde{E}_n=[n+ 1/2-3\tilde{\gamma}/(2\aleph\delta)
(n+1/2)^2]\hbar\delta\Omega  $,
as schematically shown in Fig.\ 1(b).
Interestingly, we find here that the $n^*$th level is the highest one
(noting that $n^*$ labels the resonant level-pair in the laboratory frame).
Also, the level spacing in the rotating frame is much smaller
than its counterpart in the laboratory frame.
Then, it is clear that the DDO's dynamics in this rotating frame is governed
by the interplay of $F_0 x$-induced transition and environmental dissipation.
More interestingly, in the presence of $F_0 x$, we can diagonalize
the transformed Hamiltonian $\tilde{H}_S$,
and denote the eigenstates by $\tilde{\psi}_{n}$.
It is found that a one-to-one correspondence
exists between $\psi_{n}$ and $\tilde{\psi}_{n}$,
which can be understood in the spirit of adiabatic switching.
Then, we may regard $\tilde{\psi}_{n^*}$- and $\tilde{\psi}_0$-dominated
states (wavepackets) as two {\it attractors}, which direct the evolution,
determine the final state, and are responsible to the bistable behavior.
Here we may mention that the classical dynamics of DDO can lead to
possible chaotic motion,
see, for instance, a brief discussion in Ref.\ \cite{Devoret014524}.
For the purpose of quantum detector applications, however,
the chaotic motion should be avoided by setting a low bound
of detuning \cite{Devoret014524}.
In the present work, as we did previously in Ref.\ \cite{Guo10},
we have similarly set an upper bound of detuning in order to
eliminate the possible chaotic motion.
In the mesoscopic regime,
the $\tilde{\psi}_{n^*}$- and $\tilde{\psi}_0$-dominated wavepackets,
which can be regarded as the counterparts of classical attractors,
consist of a couple of discrete states.
In the following, these ``quantized" attractors will be
named also small amplitude state (SAS) and large amplitude state (LAS).

\begin{figure}
\center
\includegraphics[scale=0.7]{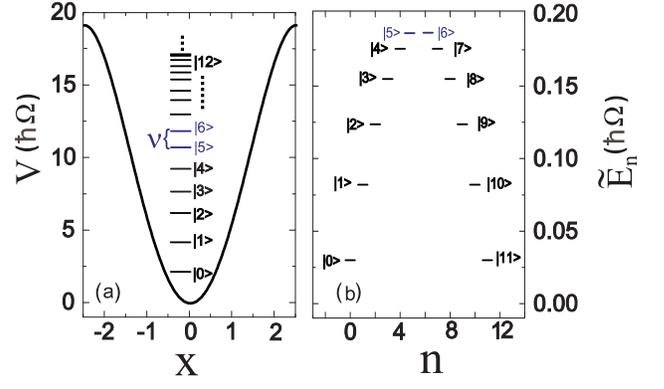}
\caption{(Color online) Energy diagram of the non-driven Duffing oscillator
in (a) the laboratory frame, and (b) the rotating frame. }
\end{figure}

\section{ Master Equation in Rotating Frame}

To account for the environmental effects on the DDO,
a Lindblad-type master equation can be employed
\cite{Katz07,Dykman042108,Dykman011101,low-lying},
simply following the typical treatment
for cavity photons (oscillators) in quantum optics.
In this work, however, viewing the ``$x$-$x_j$" type interaction
assumed by the Feynman-Vernon-Caldeira-Legget model
which is actually beyond the rotating wave approximations
\cite{Katz07,Dykman042108,Dykman011101,low-lying},
we would like to use a different type of master equation,
say, the Born-Markov-Redfield (BMR) version \cite{Tan97,Yan98,Sch98},
to study the quantum dynamics of a weakly damped DDO as it is.
Following the notation proposed in Ref.\ \cite{Yan98},
the BMR master equation for DDO's reduced density matrix reads
 \be\label{ME-LF}
 \setlength{\arraycolsep}{0.0pt}
 \dot{\rho}(t)=-i\hbar^{-1}\mathcal{L}_S\rho(t)-\hbar^{-2}\int_0^td\tau\langle
 \mathcal{L}_I(t)\mathcal{G}(t,\tau)\mathcal{L}_I(\tau)
 \mathcal{G}^\dagger(t,\tau)\rangle\rho(t).
 \ee
Here $\mathcal{L}_S(\cdots)=[H_S,(\cdots)]$ is associated
with the DDO Hamiltonian in laboratory frame, while
$\mathcal{G}(t,\tau)(\cdots)= G(t,\tau)(\cdots)G^\dagger(t,\tau)$
is the propagator in Liouvillian space.
The interaction Liouvillian  $\mathcal{L}_I$ is defined by
$\mathcal{L}_I(\cdots)=[ {H_I},(\cdots)]$,
where $H_I=-x\sum_j\lambda_j x_j\equiv -x X_E$ describes the coupling
of DDO to environment.
The average in \Eq{ME-LF} means
$\langle(\cdots)\rangle = \mathrm{tr}_E[(\cdots)\rho_E]$,
with $\rho_E$ the thermal equilibrium density operator of the environment.
In laboratory frame, the driving in $H_S$ is time-dependent,
which would complicate the dissipation terms in \Eq{ME-LF}
and make the numerical simulation difficult.
To overcome this difficulty, we transform \Eq{ME-LF} into the
rotating frame:
$ \dot{\tilde{\rho}}(t)=-i\hbar^{-1}\tilde{\mathcal{L}}_S\tilde{\rho}(t)
 -\hbar^{-2}\int_0^td\tau\langle\tilde{\mathcal{L}}_I(t)
 \tilde{\mathcal{G}}(t,\tau)\tilde{\mathcal{L}}_I(\tau)
 \tilde{\mathcal{G}}^\dagger(t,\tau)\rangle\tilde{\rho}(t)  $.
Here, the various transformed quantities are defined as:
$\tilde{\rho}(t)={U}^\dagger\rho(t) {U}$,
$ \tilde{\mathcal{L}}_S(\cdots)=[ {U}^\dagger
{H_S} {U}+i\hbar\dot{ {U}}^\dagger {U},(\cdots)]
=[{\tilde{H}_S},(\cdots)]$,
$\tilde{\mathcal{L}}_I(\cdots)=[{U}^\dagger
{H_I} {U},(\cdots)]=[ {\tilde{H}_I},(\cdots)]$,
and $\tilde{\mathcal{G}}(t,\tau)$ is the propagator
associated with ${\tilde{H}_S}$.
Note that in the rotating frame the coupling Hamiltonian
becomes time dependent, i.e.,
$ {\tilde{H}_I}
=-U^{\dagger}xUX_E=-\sqrt{\hbar/2m\Omega}
({a}^{\dagger}e^{i\nu t} + {a}e^{-i\nu t})X_E  $.
Inserting this result into the EOM of $\tilde{\rho}(t)$ yields
\begin{eqnarray}\label{ME-R2}
 \setlength{\arraycolsep}{0.0pt}
 && \dot{\tilde{\rho}}(t)=-i\hbar^{-1}\tilde{\mathcal{L}}_S\tilde{\rho}(t)
  -\frac{1}{4\aleph}\{[ {a}^{\dagger},(C(-\tilde{\mathcal{L}}_S
  +\nu) {a})\tilde{\rho}]    \nonumber\\
 &&~~~ +[ {a},(C(-\tilde{\mathcal{L}}_S-\nu)
  {a}^{\dagger})\tilde{\rho}]
  +e^{i2\nu t}[ {a}^{\dagger},(C(-\tilde{\mathcal{L}}_S-\nu)
  {a}^{\dagger})\tilde{\rho}]    \nonumber\\
 &&~~~ +e^{-i2\nu t}
  [ {a},(C(-\tilde{\mathcal{L}}_S+\nu) {a})\tilde{\rho}]
  + {\rm H.c.} \} .
 \end{eqnarray}
In deriving this result,
the well established Markov-Redfield approximation has been applied.
Accordingly, the spectral function, $C(\tilde{\mathcal{L}}_S)$,
is a Fourier transform of the environment correlator:
$ C(\tilde{\mathcal{L}}_S)=\int_{-\infty}^{+\infty}
dt C(t)e^{i\hbar^{-1}\tilde{\mathcal{L}}_St} $, and
$C(t)=\mathrm{Tr}_E[ {X_E}(t) {X_E}(0)\rho_E]$.

\Eq{ME-R2} is the desired equation we obtain in the rotating frame.
To relate it with other work, we first drop the fast oscillating terms
in \Eq{ME-R2} under the rotating wave approximation.
Then, we assume further a {\it harmonic} approximation:
$\tilde{\mathcal{L}}_S a\approx -\hbar\delta\Omega a$, and
$\tilde{\mathcal{L}}_S a^{\dagger}\approx \hbar\delta\Omega a^{\dagger}$.
As a result, the well-known Lindblad master equation is obtained:
$ \dot{\tilde{\rho}}(t)=-i\hbar^{-1}\tilde{\mathcal{L}}_S\tilde{\rho}(t)
 + \kappa \{ [1+n(\Omega)]{\cal D}[a]\tilde{\rho}
    + n(\Omega){\cal D}[a^{\dg}]\tilde{\rho}   \} $ ,
where the Lindblad superoperator is defined through
${\cal D}[A]\tilde{\rho}\equiv A\tilde{\rho}A^{\dg}
 -\frac{1}{2}\{A^{\dg}A,\tilde{\rho} \} $.
In obtaining this result,
the explicit form of the spectral function has been used,
i.e., $C(\omega)=2[1+n(\omega)]J(\omega)$,
where $n(\omega)$ is the Bose function.
We notice that in Ref.\ \cite{Dykman011101} the study of quantum activation
was actually based on this same equation, together also with a few techniques
and approximations involved in the later analysis.


\section{Quantum Shift of the Bifurcation Point}

For comparative purpose, we first outline the result
from a classical analysis.
It is well known that the classical DDO obeys \cite{Devoret014524}:
$ m\ddot{{x}}+m\Omega^2{x}+m\kappa\dot{{x}}-4\gamma{x}^3=-F(t)$.
Define dimensionless variables
$\tau=\Omega t$, $w=\nu/\Omega$,
$f=\sqrt{{4\gamma}/({m^3\Omega^6}})2F_0$,
$Q=\Omega/\kappa$, and $\Delta=-2Q\delta$;
and introduce rotating transformation
$ x(\tau)=[\tilde{x}(\tau)e^{iw\tau}/2+c.c.]/\sqrt{{4\gamma}/({m\Omega^2}})$.
Then the slowly varying amplitude $\tilde{x}(\tau)$ of the DDO
in the rotating frame satisfies the following EOM \cite{Devoret014524}:
$ 2i\dot{\tilde{x}}(\tau)=[(\Delta-i)/Q
  +3|\tilde{x}|^2 /4 ]\tilde{x}-f $.
Stationary solution of this equation yields a
bistability diagram as shown in Fig.\ 2(a).
Moreover, two critical driving strengths,
$f_B(\Delta)$ and $f_{\bar{B}}(\Delta)$,
can be obtained in the limit $Q\gg1$:
$f_{B,\bar{B}}(\Delta/\Delta_c)=f_c /(2x^{3/2})
\sqrt{1+3x^2\pm(1-x^2)^{3/2}}|_{x=\Delta_c/\Delta}  $.
Here $\Delta_c=-\sqrt{3}$, and $f_c=2^{5/2}/(3^{5/4}\sqrt{\Delta^3})$.
Accordingly, after restoring dimensional units, the critical driving strengths
read $F_{B,\bar{B}}=\sqrt{m^3\Omega^6/(16\gamma)}f_{B,\bar{B}}$.
In what follows, we will focus on the most important upper
point $F_B$, which is to be re-denoted as $F_c$
and taken as the unit of the driving force.

Unfortunately, in {\it mesoscopic} regime, as we will see later,
the critical driving strength $F_c$ determined above does not
match the result from numerical simulation.
It would thus be desirable to develop a {\it quantum}
version for the EOM of $\tilde{x}(\tau)$,
in order to reach consensus with the direct numerical simulation.
Following Ref.\ \cite{Joo03}, given that the reduced density matrix
satisfies the Lindblad master equation, in Heisenberg picture
the operator $a$ should obey an equation of motion as follows
\bea\label{a-eq}
\dot{a}(t)=-i\hbar^{-1}[a,\tilde{H}_S] +\kappa\bar{\cal D}[a]a
    = -i\hbar^{-1}[a,\tilde{H}_S] - \frac{\kappa}{2} a   ,
\eea
where the {\it dual}-Lindblad superoperator is defined through
$\bar{\cal D}[A]a \equiv A^{\dg}aA -\frac{1}{2}\{A^{\dg}A, a \} $.
Here we assumed the low temperature limit $n(\Omega)\ll 1$.
Moreover, below $F_c$ and starting with a small-amplitude state,
the subsequent evolution will largely remain in a coherent state
\cite{Kni05,Khe99,note-3}.
Then, in coherent state representation and
relating the coherence number $\alpha(\tau)$
with a complex amplitude
$\tilde{x}(\tau)=\sqrt{8\gamma/(m\Omega^2)}\alpha(\tau)/\sqrt{n}$,
from \Eq{a-eq} we obtain the same EOM for $\tilde{x}(\tau)$ as above,
but with a quantum mechanically shifted detuning
\begin{equation}\label{Q-shift}
\tilde{\Delta}=-2Q(\delta-3\tilde{\gamma}/\aleph) ,
\end{equation}
instead of the {\it classical} result $\Delta=-2Q\delta$.
In classical case, $\aleph$ is large,
e.g., $\aleph \simeq 366$ as we estimated
from the experimental circuit parameters,
which makes the quantum shift negligibly small.
Nevertheless, in mesoscopic regime, e.g., $\aleph \simeq 12$ in present case,
this quantum shift cannot be neglected,
as shown in Fig.\ 2(b) and (c), where the critical driving strength
moves to $0.77F_c$.
Below we illustrate this quantum shift in phase space in terms of
the Wigner function.

\begin{figure}
\center
\includegraphics[scale=0.7]{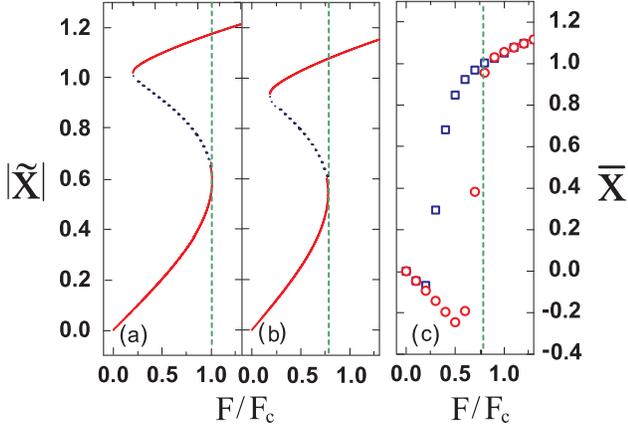}
\caption{(Color online) (a) and (b):
Bifurcation diagram of the DDO based on the EOM of $\tilde{x}(\tau)$,
(a) without and (b) with accounting for the quantum shift
of the effective detuning [see \Eq{Q-shift}].
The dashed vertical lines indicate the respective bifurcation points.
(c): Results from numerical simulation of
$\bar{x}(t)$ =$\mathrm{Tr}[{x}\tilde{\rho}(t)]$ based on \Eq{ME-R2}.
We choose the $\bar{x}(t)$ at $t=160*(2\pi/\Omega)$
to represent the steady state amplitude $\bar{x}$.
The circles and squares stand for results from different initial
conditions (i.e. the SAS and LAS), from which we observe the
hysteresis behavior.
Here, again, the dashed vertical line indicates the bifurcation point
in (b), say, $0.77F_c$, which agrees well with the numerical simulation.
Parameters: $\kappa=0.01$, $T$=5mK, $\delta=0.065$, and $\aleph=12$.   }
\end{figure}

\begin{figure}
 \center
 \includegraphics[scale=0.8]{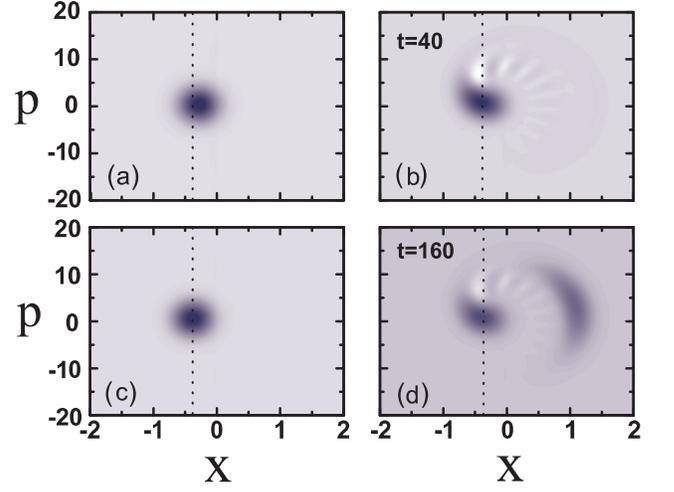} \caption{ (Color online) (a) and (b): Wigner
function of the coherent state determined from the steady-state
solution of the EOM of $\tilde{x}(\tau)$, (a) without and (b) with
accounting for the quantum shift. (c) and (d): Wigner function
from numerical simulation based on \Eq{ME-R2}, at
$t=40*(2\pi/\Omega)$ and $160*(2\pi/\Omega)$, respectively. 
The dashed vertical line in each sub-figure indicates the valid ``$x$" center
of Wigner function of the SAS.
Parameters:
$\kappa=0.01$, $T$=5mK, $\delta=0.065$, $\aleph=12$, and $F_0=0.7F_c$. }
\end{figure}

The Wigner function is defined as:
$ W(x,p,t)=1/(\pi\hbar)\int_{-\infty}^{+\infty}
\langle x+x'|\rho(t)|x-x'\rangle\exp(-i2px'/\hbar)dx'$.
In Fig.\ 3(a) and (b), for comparative purpose, we plot the Wigner function
simply using $\tilde{\rho}_s\equiv|\alpha\rangle\langle\alpha|$,
where the coherence number $\alpha$ is determined from $\tilde{x}$,
based on the steady-state solution of the amplitude EOM without and with
accounting for the quantum shift,
while in Fig.\ 3(c) and (d) we show the results from direct simulation.
We see that in mesoscopic regime it is essential to account for the
quantum shift, in order to make the amplitude EOM agree with
numerical simulation, as indicated by the dashed vertical lines in Fig.\ 3.
(The vertical line in each sub-figure indicates the valid ``$x$" center
of Wigner function of the SAS.) It is then observed that the classical
result in Fig.\ 3(a) has considerable deviation.
Moreover, from Fig.\ 3(d) we get an insight that starting on from
certain transient stage, the oscillator evolves into a mixed state,
which can be formally expressed as:
$ W(x,p,t)=P_S(t)W_S(x,p)+P_L(t) W_L(x,p,t) $.
Here, $W_S(x,p)$ and $W_L(x,p,t)$ are the Wigner
functions of the intrinsic SAS and LAS, say, the two attractors,
while $P_S(t)$ and $P_L(t)$ are the respective occupation probabilities.

\begin{figure}
 \center
 \includegraphics[scale=1.15]{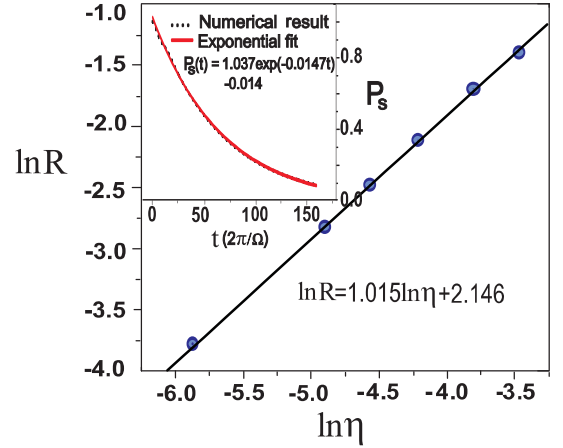}
 \caption{(Color online)
Scaling behavior of the dynamic quantum tunneling rate
with the driving distance to the shifted bifurcation point,
say, $\eta =(0.77F_c)^2-F_0^2$.
Here, the circles are result from numerical
simulation, while the linear fit gives $\alpha=1.015$
for the tunneling action $R \varpropto\eta^{\alpha}$.
Inset: an illustrative example of exponential fitting
for the occupation probability of the SAS, under driving $F_0=0.76F_c$.
Parameters:
$\kappa=0.01$, $T$=5mK, $\delta=0.065$, and $\aleph=12$. }
\end{figure}

\section{Tunneling Rate and Scaling Exponent}

Based on the structure of the mixed state $W(x,p,t)$,
we can formulate a way to determine
the dynamic tunneling rate from SAS to LAS as follows.
Originally, in laboratory frame, this rate can be extracted from the
occupation probability of the SAS via $\Gamma_t=-dP_S(t)/dt$, while
$P_S(t)=\langle\alpha e^{i\nu t}|\rho(t)|\alpha e^{i\nu t}\rangle$.
Note that here we have used the fact that the SAS in laboratory frame is
a {\it rotating} coherent state.
More conveniently, transformed to the rotating frame,
$P_S(t)=\mathrm{Tr}[\tilde{\rho}_s\tilde{\rho}(t)]$,
where
$\tilde{\rho}_s\equiv U^{\dagger}(t)\rho_s(t) U(t)
=|\alpha\rangle\langle\alpha|$ is static,
and $\tilde{\rho}(t)=U^{\dagger}(t)\rho(t) U(t)$
is the DDO state in the rotating frame described by \Eq{ME-R2}.

In the inset of Fig.\ 4 we show a representative $P_S(t)$, obtained from
numerical simulation using \Eq{ME-R2} and the formalism outlined above,
from which we see that a single exponential fit can well characterize $P_S(t)$.
This indicates nothing but a rate process from SAS to LAS.
Moreover, the success of a single exponential fit indicates a dominant
forward process from SAS to LAS. That is, at the early escape stage,
the backward process from LAS to SAS has not yet happened.
This is similar to the situation in determining the tunneling rate
in double-well problem, where the backward tunneling process is
negligibly small at certain early time stage.

Therefore, using this numerical approach we can extract the dynamic
tunneling rate ($\Gamma_t$) from SAS to LAS. Further,
following Ref.\ \cite{Dykman011101}, we assume $\Gamma_t=Ce^{-R/\lambda}$.
In this proposed form, $C$ is an irrelevant prefactor, while the exponential
factor $e^{-R/\lambda}$ originates from an {\it effective} activation process.
In limiting cases, such as for classical thermal activation,
$R$ is the activation energy and $\lambda$ the temperature;
while for quantum tunneling through a barrier, $R$ is
the tunneling action and $\lambda$ the Plank constant.
In our present case, it is a generalization. We may view $R$
as an effective activation energy and $\lambda$ an effective
Planck constant or temperature.

Very interestingly, it was found in Ref.\ \cite{Dykman011101} that
the dynamic tunneling action $R$ displays a perfect
scaling behavior with the driving distance to the critical point,
which is defined as $\eta = F_c^2-F^2_0$.
Quantitatively, it was found $R\varpropto\eta^{\alpha}$ and $\alpha=3/2$.
Here, for the mesoscopic DDO, owing to the quantum shift
of the critical point, which moves to $0.77F_c$ under the assumed
parameter condition as shown in Fig.\ 2,
we define accordingly the driving distance as $\eta = (0.77F_c)^2-F^2_0$.
Remarkably, we demonstrate in Fig.\ 4 that the scaling behavior
of $R\varpropto\eta^{\alpha}$ still holds, yet with an alternative scaling
exponent, $\alpha\simeq 1$, instead of $\alpha=3/2$ as found by Dykman
\cite{Dykman011101}.

We noticed that in Ref.\ \cite{Sta06}, scaling behavior
of the transition rate with the driving {\it frequency} (but not the
driving {\it strength}) was analyzed to give $\alpha\simeq 1.3\sim 1.4$,
by a rough fitting from a few experimental data.
Meanwhile, in the experiments by Siddiqi {\it et al.} \cite{Sid0305},
an effective potential with a barrier height scaled as
$\Delta U^0_{dyn}\propto\left[1-(F_0/F_c)^2\right]^{3/2}$,
was employed to analyze their measured data.
By means of a thermal-activation rate
$\propto \exp\left(-\Delta U^0_{dyn}/k_BT\right) $,
this would result in a scaling exponent $\alpha=3/2$,
while a rough quantum WKB estimate for the quantum tunneling rate
$\propto \exp\left(-\sqrt{\Delta U^0_{dyn}}a/\hbar\right)$,
where $a$ is an effective width of the barrier,
is seemingly to give $\alpha=3/4$.

Moreover, in Ref.\ \cite{Dyk05}, it was shown that both the exponents
``3/2" and ``1" can be obtained, respectively, resulting from
a {\it local} and {\it nonlocal} Hamiltonian bifurcation.
Nevertheless, we should distinguish Ref.\ \cite{Dyk05} from our present
study in at least two aspects:
{\it (i)}
The rates obtained in Ref.\ \cite{Dyk05} are classical ones,
i.e., the thermal-activation-caused escape rates, since they are
respectively related to an ``effective potential" that is then
inserted into the thermal-activation rate formula.
{\it (ii)}
For the DDO model under our present study, the ``local bifurcation" is near
the larger bifurcation point, and the ``nonlocal bifurcation" is close to
the smaller one. This clarification is clearly stated in Sec VII of
Ref.\ \cite{Dyk05}. Our numerical study, however, is focused on the region
near the larger bifurcation point, thus should correspond to the
``local bifurcation", and should be compared with the exponent ``3/2".

In a recent work\cite{Guo10},
we performed a real time simulation for the quantum dynamics
of the mesoscopic DDO in laboratory frame,
in which the driving field was not taken into account in the dissipation terms.
Also, in Ref.\ \cite{Guo10}, the quantum shift was canceled owing to
the extra term ``$x^2\sum_i\lambda_i^2/(2m_i\omega_i^2)$"
in the Caldeira-Legget model.
In the present work, however, absorbing this term into the system Hamiltonian,
which effectively renormalizes the DDO's frequency,
makes the quantum shift evident.
In regard to the scaling behavior, quite desirably,
both studies of Ref.\ \cite{Guo10} and the present one
give consistent result. This strongly implies that the scaling exponent
$\alpha=3/2$ is non-universal, and $\alpha = 1$ is an alternate
scaling exponent in the mesoscoic regime.
Further investigation for the crossover behavior from a mesoscopic
to the usual classical regime is of particular interest.
However, such kind of study will encounter increasing difficulty
in real time simulation, which is to be the task of our forthcoming research.

\section{Summary}

To summarize, in a mesoscopic regime we investigated the dynamic quantum
tunneling of the driven Duffing oscillator.
Owing to the mesoscopic nature, we found that the critical driving strength
has a quantum shift, and the tunneling action (extracted exponentially from
the tunneling rate) exhibits a perfect linear scaling behavior with
the driving distance to the quantum shifted critical point.

\vspace{1cm} {\it Acknowledgements.---} 
This work was supported by the National Natural Science Foundation
of China, the 973 project, the Fundamental Research Fund for
Central Universities, and the Fund for Doctoral Training.



\begin{thebibliography}{27}
\bibitem{note-3}
That the SAS is an approximate coherent state can be understood
as follows. For a dissipative harmonic oscillator
described by a Lindblad master equation, e.g., a damping optical
cavity under laser driving, the stationary state is exactly a coherent state.
An elegant proving for this can be found in the book \cite{Kni05}.
For the DDO problem, since the SAS is near the bottom of the potential
well where the anharmonicity is negligible,
we can then expect that the SAS is an approximate coherent state.
\end{thebibliography}
\end{document}